\documentclass[12pt,a4paper]{article}

\usepackage{ifpdf,amsmath,amssymb,amsthm}
\usepackage{fullpage}
\usepackage{graphicx}
\ifpdf
\usepackage{cmap}
\usepackage[pdftex,colorlinks=true,linkcolor=blue,unicode=true,hyperfootnotes=false]{hyperref}
\fi

\renewcommand{\ge}{\geqslant}
\begin{document}

\title{Algorithmic randomness and splitting of supermartingales}
\author{Andrej Muchnik\thanks{%
  This work was done by Andrej Muchnik (1958--2007) in 2003.
  Soon after that, A.\,Chernov wrote a draft version of the paper
  based on a talk of Andrej Muchnik. Then An.\,Muchnik looked it
  through and planned to edit the text, but did not manage to do
  it. The current text is prepared by A.\,Chernov and A.\,Shen
  in 2007--2008; they are responsible for any possible errors and
  inaccuracies. Supported by RFBR and CNRS ANR grants.}
}

\date{}

\maketitle

\begin{abstract}
Randomness in the sense of Martin-L\"of can be defined in terms
of lower semicomputable supermartingales. We show that such a
supermartingale cannot be replaced by a pair of supermartingales
that bet only on the even bits (the first one) and on the odd
bits (the second one) knowing all preceding bits.
\end{abstract}

\section{Randomness and lower semicomputable supermartingales}

The notion of algorithmic randomness (Martin-L\"of randomness)
for an infinite sequence of zeros and ones (with respect to
uniform Bernoulli distribution and independent trials) can be
defined using supermartingales. In this context, a
\emph{supermartingale} is a non-negative real-valued function
$m$ on binary strings such that
        $$
m(x) \ge \frac{m(x0)+m(x1)}{2}
        $$
for all strings~$x$. Any supermartingale corresponds to a
strategy in the following game. In the beginning we have some
initial capital ($m(\Lambda)$, where $\Lambda$ is the empty
string). Before each round, we put part of the money on zero,
some other part on one, and throw away the rest. Then the next
random bit of the sequence is generated, the correct stake is
doubled and the incorrect one is lost. In these terms, $m(x)$ is
the capital after bit string~$x$ appears. (If the option to
throw away a portion of money is not used, then the inequality
becomes an equality, and the function $m$ is a
\emph{martingale}.)

We say that a supermartingale $m$ \emph{wins} on an infinite
sequence~$\omega$ if the values of $m$ on the initial segments
of $\omega$ are unbounded. The set of all sequences where $m$
wins is called its winning set.

Algorithmic probability theory often uses \emph{lower
semicomputable} supermartingales. This means that for each $x$
the value $m(x)$ is a limit of a non-decreasing sequence of
non-negative rational numbers $M(x,0), M(x,1),\ldots$, and the
mapping $(x,n)\mapsto M(x,n)$ is computable.

The class of all lower semicomputable supermartingales has a
maximal element (up to a constant factor). Its winning set
contains winning sets of all lower semicomputable
supermartingales; this set is called the set of \emph{nonrandom}
sequences. The complement of this set is the set of
\emph{random} sequences.

This definition is equivalent to the standard definition given
by Martin-L\"of (e.\,g., see~\cite{uss}); sometimes it is called
a criterion of Martin-L\"of randomness in terms of supermartingales.

Some authors consider also a larger class of ``computably
random'' sequences that one gets if lower semicomputable
supermartingales are replaced by computable martingales (for
simplicity, one can take rational-valued martingales). There is
no maximal computable martingale, and a sequence is called
computably random if \textsl{any} computable martingale is
bounded on its initial segments. However, we keep Martin-L\"of
definition as the main one; in the rest of the paper ``random''
without additional adjectives means ``Martin-L\"of random''.

\section{Odd/even decomposition of martingales}

One can observe that for computable martingales it is sufficient
to consider separately bets on odd steps only and bets on even
steps only to get the definition of randomness. We say that a
(super)martingale \emph{bets only on even steps} (or, in other
words, does not bet on odd steps) if $m(x)=m(x0)=m(x1)$ for all
strings $x$ of even length. (For instance, a martingale is not
betting at the third step if after the third coin tossing the
capital does not change, that is, if $m(x0)=m(x1)=m(x)$ for
every~$x$ of length~$2$. In terms of the game this means that
one half of the capital is put on zero and one half is put on
one, and in any case the capital remains the same.) Similarly, a
(super)martingale \emph{bets only on odd steps} if
$m(x0)=m(x1)=m(x)$ for every~$x$ of odd length.

Now let us give the precise formulation of the observation
above: \emph{for any computable martingale $t$ there exist two
computable martingales $t_0$ and $t_1$ such that $t_0$ bets only
on even steps and $t_1$ bets only on odd steps, and the
following holds: if $t$ wins on some sequence $\omega$, then
either $t_0$ or $t_1$ wins on $\omega$.}

This implies that the winning set of $t$ is included in the
union of the winning sets of $t_0$ and~$t_1$.

\textbf{Proof}: Adding a constant if necessary, we may assume
that $t$ is strictly positive everywhere. Then we construct two
martingales $t_0$ and $t_1$ as follows: on even steps, $t_0$
divides its capital between zero and one in the same proportion
as $t$ does, and $t_1$ does not bet (puts equal stakes on zero
and one); on odd steps, $t_0$ and $t_1$ change roles. Then the
gain of $t$ (the current capital divided by the initial capital)
equals to the product of the gains of $t_0$ and $t_1$.
Therefore, if both $t_0$ and $t_1$ are bounded on prefixes of
some sequence, so is $t$.

\medskip

In other words, defining randomness with respect to computable
martingales, we may restrict ourselves to martingales that bet
every other step (on even or odd steps only). A similar
statement is true for the Mises -- Church definition of
randomness (that uses selection rules, see~\cite{uss}): it is
enough to split a selection rule into two rules; one selects
only even terms, the other selects only odd terms.

It turns out, however, that for lower semicomputable
supermartingales (and Martin-L\"of random sequences) the
analogous statement is wrong, and this is the main result of the
paper.

\medskip

\textbf{Theorem}.
\emph{There exists a Martin-L\"of non-random sequence $\omega$
such that no supermartingale betting every other step \textup(on
even steps or on odd steps\textup) wins on $\omega$.}

\smallskip

This result will be proved in the next sections. Now let us
point out its relations to the van~Lam\-bal\-gen theorem on
randomness of pairs.

Any bit sequence~$\alpha$ can be splitted into two subsequences
of even and odd terms $\alpha_0$ and $\alpha_1$. It is easy to
see that for a (Martin-L\"of) random sequence $\alpha$, both
sequences $\alpha_0$ and $\alpha_1$ are random. However, the
converse statement is not true (trivial counterexample: even if
$\alpha_0=\alpha_1$ is random, the sequence with doubled bits is
not).

A well-known theorem of M.\,van Lambalgen~\cite{lam} gives a
necessary and sufficient condition for the randomness of
$\alpha$: it is random if and only if $\alpha_0$ is random with
the $\alpha_1$-oracle and $\alpha_1$ is random with the
$\alpha_0$-oracle. (Informally, this means that even the player
that can see any elements of $\alpha_1$ before betting on
$\alpha_0$ and vice versa cannot win on any of these two
sequences.) It would be natural to conjecture that there is no
need to ``look ahead'' (choosing the stakes on the $(2n-1)$th
step, one does not need to use the bits at positions $2n$,
$2n+2$, $2n+4$, etc.). Surprisingly, this conjecture is wrong,
as the theorem shows.

This can be considered as a paradox: imagine two referrees who
toss a coin during some tournament alternatively (for even and
odd days respectively); we would expect that if each of them
does her job well (her subsequence is ``on-line random'' in the
context of the entire sequence, see~\cite{online}), then the
entire sequence is random. The example constructed in this paper
show that this is not the case if we define randomness in terms
of supermartingales: it may happen the the entire sequence is
flaw but the flaw cannot be attributed to one of the referees.

\section{Construction of a game}

We present the proof using a certain infinite game of two
players called Mathematician (M) and Adversary (A).

Adversary constructs (enumerates from below) two
supermartingales $t_0$ and $t_1$ (that make bets on even and odd
steps, respectively). Mathematician constructs a supermartingale
$t$. At each move, the players provide some approximations from
below for their supermartingales; these approximations have only
finitely many non-zero values and are itself supermartingales
(betting on admissible steps only, for A's supermartingales).
The sequence of approximating supermartingales is non-decreasing
(the players cannot decrease the values already announced). Thus
the moves of the players are constructive objects (can be
encoded by binary strings, etc.). The players move in turn and
can see moves of each other (so we have a perfect information
game). Additionally, the values of all the supermartingales on
the empty string are required to be $1$ (this normalization
prevents infinite limit value on any string).

The game is an infinite sequence of moves of the players; it
determines three limit supermartingales. M wins if there exists
a sequence $\omega$ such that M's supermartingale $t$ wins on
$\omega$ (is unbounded on the prefixes of $\omega$) and both A's
supermartingales $t_0$ and $t_1$ do not win (are bounded on the
prefixes of $\omega$).

\textbf{Main Lemma}. \emph{M has a computable winning
strategy in the game.}

It is easy to see that this lemma implies the theorem. Actually,
the standard argument (see~\cite{uss}) shows that there are two
maximal (up to a constant factor) lower semicomputable
supermartingales $t_0$ and $t_1$ of the type considered (making
bets every other step). Let us run M's winning strategy against
the enumeration of these two supermartingales by A. (Note that
in this case A does not use the information about M's moves.)

Since the strategies of both players are computable, the limit
supermartingale $t$ is lower semicomputable. Since M's strategy
wins, there is a sequence $\omega$ such that $t$ wins on
$\omega$ but both $t_0$, $t_1$ (and therefore any other lower
semicomputable supermartingale of the type considered) do not
win on $\omega$. The theorem follows.

It remains to prove the Lemma. In the next sections, we redefine
this game as a game on an infinite binary tree and define
versions of this game on a finite tree. Then we explain how to
combine winning strategies for the games on finite trees into
one winning strategy for the game on the infinite tree. Finally,
we describe a winning strategy on a finite tree.

\section{Games on finite and infinite trees}

It is convenient to consider an infinite binary tree that is a
``field'' of the game described. The nodes of the tree are
binary strings ($x0$ and $x1$ are children of $x$). The players
increase the current values (approximations from below) of their
supermartingales (one value per node for each supermartingale).
``After the game ends'' (the quotes are used since the game is
infinite), a referee comes and looks for an infinite tree branch
with the properties described in the winning conditions above.

Let us define an auxiliary game on a finite binary tree of a
certain height~$h$. As before, the players make their moves and
increase the current values of their supermartingales ($t$ for M
and $t_0$, $t_1$ for A). The supermartingales $t_0$ and $t_1$
satisfy the same restrictions as before: $t_0$ bets on even
steps, $t_1$ bets on odd steps. The values of all three
supermartingales in the root (on the empty word) are equal to
one all the time. In the other tree nodes, the supermartingales
are equal to zero in the beginning and then the players increase
them step by step.

The game on a finite tree is still infinite. ``At the end'' of
this infinite game, we consider the limit values of the
supermartingales at the leaves of the tree to determine the
winner. The winning condition will be quite technical, but the
idea is as follows: M wins if there exists a leaf where M's
supermartingale $t$ is substantially greater than~$1$, while on
the entire path to this leaf both A's supermartingales $t_0$ and
$t_1$ do not exceed~$1$ significantly
(actually, this is a ``finite version''
of the winning condition on the infinite tree).

Let us assume for a while (later we will need to consider a
more complicated game) that M has a winning strategy for the
following version of the game: $M$ can guarantee that in one of
the leaves $t$ is greater than~$2$ while $t_0$ and $t_1$ do not
exceed $1$ on the path to this leaf.

Then this winning strategy for M can be used as a building block
for M's strategy on the infinite tree. Indeed, we can start the
second game on another finite tree whose root is the winning
leaf in the first tree, but with twice more money. (This means
that the actual M's moves on the subtree of the infinite tree
are twice bigger that M's moves in the second finite game. In
other words, we apply the winning strategy for the finite tree
doubling all values of $t$.) In the limit, this second game
gives a leaf (of the second finite tree) where M's
supermartingale $t$ exceeds $4=2^2$ and A's supermartingales
$t_0$, $t_1$ still do not exceed~$1$ (on the whole path from the
root of the infinite tree). One more copy of the game is then
started from this leaf (with factor~$4$), it gives $8=2^3$ in
some leaf etc. In the limit (where infinitely many finite games
are combined), we get an infinite branch where $t$ goes to
infinity while $t_0$ and $t_1$ are bounded.

This description is oversimpified and ignores some important
points. First of all, we should keep in mind that the game on
the finite tree is infinite, and thus M must start the next
round (on the second finite tree grown from the leaf of the
first one) when the first game is still in progress (and
therefore, the leaf chosen may be discredited later when values
of $t_0$ and $t_1$ increase).

From the formulation of the winning condition on the
finite tree, we see that the condition (for a certain leaf) can
be fulfilled at some moment ($t$ is large whereas $t_0$, $t_1$
are small yet), and then be violated (when A increases $t_0$ or
$t_1$). Note that after that the winning condition cannot be
fulfilled again, hence a leaf cannot become a winning candidate
for the second time.

\begin{figure}[h]
\begin{center}
\includegraphics[scale=1.0]{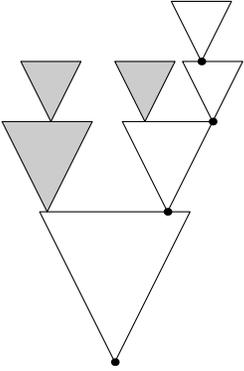}
\end{center}
\caption{Active and discontinued games on finite trees}\label{supermart1}
\end{figure}

Starting a new subtree from the current winning candidate (and
discarding this subtree when and if this leaf is discredited),
we get a picture like Figure~\ref{supermart1}.
The grey triangles represent subtrees where the game was started
and then cancelled (the candidate in their root was discredited);
the white triangles represent active (currently) games. The dots
represent current winning candidates. (We may allow several winning
candidates in the same tree and start new subtree for all of
them; however, in the picture above this is not shown. In fact,
we postpone new games and add the next tree of the
same level only when the previous one has been discarded.)

Note also that in reality M should play against A on the
infinite tree (and not against many adversaries on finite
trees). However, the moves of A on the infinite tree can be
easily translated into moves of the adversaries in the active
games on finite trees.

By the definition of the winning strategy on the finite tree,
M's supermatingale $t$ doubles at each ``level'' compared to its
value at the previous ``level''; on the other hand, A's
supermartingales $t_0$, $t_1$ do not exceed~$1$ along the entire
path through the winning leaves.

Let us see what happens with this picture in the limit. At the
first level the winning candidate may change only finitely many
times. Hence, at some moment some candidate will be ``final''
and will never be discredited. The game in the subtree starting
from this candidate will never be discarded and a winning leaf
should appear there (and stabilize after finitely many changes),
and so on. In the limit, we get an infinite branch where M's
supermartingale $t$ is unbounded while both A's supermartingales
$t_0$, $t_1$ are bounded.

\section{Game on a finite tree}

\subsection*{Informal discussion}

We begin with some informal discussion, which may help to
understand the idea behind the winning strategy for a finite
tree game. However, the reader can safely proceed to the formal
statement and proof (that uses a slightly different and more
simple, but less intuitive, argument)
if this informal introduction seems unclear.

Let us make several trivial observations. M wants to make the
supermartingale $t$ greater than~$1$ in some leaf. But this
means that $t$ must remain less than~$1$ in some other leaves
(since the average, the root value, is~$1$). So what M really
needs to do is to find leaves that are not ``promising'' and not
to waste $t$ on them, saving the money for the other leaves. In
particular, M can skip the leaves that have already been
discredited ($t_0$ or $t_1$ exceeds $1$ on the path to these
leaves). If we put some $C>1$ in all the leaves except for
some ``discredited'' one, then we are done: there exists some path
on which both A's supermartingales are bounded by~$1$
(indeed, at each vertex one of the supermartnigales
does not play and the other loses in some direction)
and this
path ends in a non-discredited leaf where we have put $C$.
(To keep the average in the root not exceeding~$1$, we let
$C=N/(N-1)$ where $N$ is the number of leaves.)

So, let us try to construct a strategy for M assuming
additionally that A avoids discrediting leaves ``in advance''. M
starts by placing some $C>1$ in the leftmost leaf $x$.
\begin{figure}[h]
\begin{center}
\includegraphics[scale=1.0]{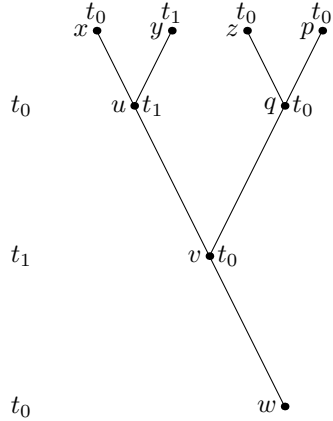}
\end{center}
\caption{The start of the game. The left column indicates
which of the supermartingales plays at a given level.}\label{supermart5}
\end{figure}
Then A
must discredit $x$ by making one of $t_0$, $t_1$ greater
than~$1$ in~$x$ or on the path to~$x$. Note that if $t_0$ or $t_1$ is
greater than~$1$ \textsl{below}~$x$ then some other leaves
become discredited at the same time, and this should not happen
according to our assumption. So A has to increase one of its
supermartingales in $x$ itself. Moreover, A has to increase the
supermartingale that bets in~$x$'s parent $u$ (say, $t_0$):
the increase of the
other supermartingale will mean its increase in $u$ and
therefore also on the way to $x$'s brother $y$,
so some leaf would become discredited before
M spent money on it (and this should not happen according to our
assumption).

At the second move, M places $C$ in $y$.
Now A cannot increase $t_0$ on the way to $y$,
because then two more leaves ($x$'s
cousins $z$ and $p$) would be discredited prematurely.
Indeed, if $t_0$ exceeds $1$ on the way to $y$, then
(due to averaging in $u$; note that $t_0$ exceeds $1$ in~$x$)
it exceeds $1$ in $u$ and
also in its father $v$, since $t_0$ does not play in~$v$, which
makes other two grandsons of $v$ ($z$ and $p$ in
the picture) discredited. Therefore, at some point
$t_1$ exceeds $1$ in $y$ (or on the way to $y$) and therefore
in $u$ (or on the way to $u$).

Then we place $C$ in the third (from left to right) leaf $z$ and
wait until A makes one of the supermartingales $t_0$ and $t_1$
greater than $1$ in $z$ or on the way to $z$. It could be only
$t_0$ for the same reasons as above (otherwise $z$'s brother
$p$ would be
discredited). Next step is to place $C$ in $p$ and wait until
one of $t_0$ and $t_1$ exceeds~$1$ in $p$ or on the
way to $p$. This could happen with $t_0$ only: indeed,
if $t_1$ exceeds $1$ on the way to $p$, then $t_1$
exceeds $1$ in both $u$ and $q$ ($p$'s father) and therefore in $v$
(averaging) and $w$ ($v$'s father where $t_1$ does not play), which
discredits four next leaves.

Continuing this process, we see that A is always forced to
follow a certain pattern, if A wants to avoid spreading down the
values of $t_0$ and $t_1$ and hence discrediting some leaves too
early. To understand this pattern, we can imagine that each
leaf contains a Boolean variable: its value ($0$ or $1$) says
which of $t_0$ and $t_1$ exceeds~$1$ in this leaf. Note that
the values of these variables spread down by simple AND-OR rules:
if the supermartingale bets in the node and exceeds~$1$ in
\textsl{both} children of this node, it should exceed~$1$ in the
node itself; the same is true if the supermartingale does not
bet in this node and exceeds~$1$ in \textsl{at least one} of
its children.

So the process resembles the evaluating of an AND-OR-tree whose
leaves carry Boolean values and AND/OR-layers alternate. ``Early
discrediting'' corresponds to the so called shortcut evaluation:
unused variables correspond to the prematurely discredited nodes.
It is easy to see that if A assigns values to the leaves of an
AND/OR-tree (say, from left to right) and wants to avoid
shortcut evaluation, there is only one way of doing so (except
for the last variable). The same is true in the game: if A wants
at any rate to avoid the cases where some leaves are discredited
before M puts some value there, and M goes from left to right, A
has essentially no choice (except for the last moment).

The idea of the actual strategy of M is to use either the advantage
that M can gain when A deviates from the pattern (and discredits some
leaf prematurely) or to use the advantage that M can gain when A
follows the pattern for the first $1/4$ of all leaves. For the
first case, when M does not need to place anything in some
prematurely discredited leaf M can safely place $C=N/(N-1)$ in
all the other leaves (where $N$ is the number of leaves). Note
that it is much less than $2$ that we planned initially but still
enough for our purposes as we see later.
For the second case,
let us consider the Boolean values
in the leaves that prevent shortcuts, i.e., the sequence
$0100010101000100...$\footnote{
    For more information about this sequence see
    http://www.research.att.com/~njas/sequences/A035263 and
    http://www.research.att.com/~njas/sequences/A096268.}
Every second place is occupied by a zero, that is, at every
second leaf, $t_0$ is greater than~$1$. Hence, two levels below,
$t_0$ is greater than~$1/2$ in all the nodes. And it is easy to
see that the sequence of labels is the same there: at every
second node, $t_0$ is greater than~$1$ again. Therefore, four
levels below the leaves, $t_0$ is greater than~$3/4$ in all the
nodes, and so on. The same works for $t_1$, if we go through the
``odd'' levels (one, three, five,... levels below the leaves).
Thus M can take any node, a sufficiently large subtree above it,
and provide that, say, $t_1$ is close to~$1$ (though, may be,
less than~$1$) in this node. Let the node be the leftmost
grandchild of the root (where $t_1$ bets, but does not bet in
the children of the root). Then $t_1$ is close to~$1$ in the
left child of the root. Therefore it cannot exceed $1$
significantly in the right child of the root, while $t_0$ does
not exceed $1$ there (since $t_0$ does not bet in the root). Now
M concentrates on the right half of the subtree and can put,
e.g., $4/3$ in all its leaves (recall that $1/4$ of all leaves
remain free and the other $1/4$ carries only $C$). But A's
supermartingales do not exceed $1$ significantly in the right child
of the root and therefore the same is true for some path in the
right half of the tree.

In both cases, M has achieved something, but these achievements
differ. To reflect this, we have to change the definition of the
finite game allowing two winning conditions for A. Let us now
proceed with the details. (The motivation as explained above is
not explicitly used in the sequel.)

\subsection*{Game on the finite tree: precise formulation}

As we have explained, we do not achieve the earlier announced
goal (M's supermartingale exceeds~$2$ in some leaf whereas A's
supermartingales do not exceed~$1$ on the path to it). Let us
explain what we can really achieve and how to use this for the
winning strategy in the infinite game.

\subsubsection*{Admissible increase of the supermartingales}

We are not able to guarantee that (for the game on the finite
tree) there exists a leaf where $t$ increases while both $t_0$,
$t_1$ still do not exceed~$1$. Instead, we have to allow some
small increase for $t_0$ and $t_1$, limited by factor
$(1+\delta)$. These coefficients are multiplied along the tree
chain, but we manage to have decreasing values of $\delta$ along
the chain so the product is bounded by some constant.

On the other hand, we cannot guarantee also that $t$ is
multiplied at least by $2$; only some smaller factor (depending
on $\delta$, as we will see) can be achieved. We have to ensure
here that the product of these factors diverges to infinity.

More formally, for every finite subtree, a special version of
lemma is used with appropriate factors (guaranteed increase for
$t$ and allowed increase for $t_0$ and $t_1$). The products of
the corresponding factors for preceding subtrees (along the path
from the root) are used as scaling factors for the current
subtree. Note that scaling factors for M and A are different and
are multiplied separately. The winning condition for the
preceding games guarantees that the actual moves of A (divided
by A's scaling factor) do not violate the rules of the game on
the finite tree, and the moves of M in the game on the finite
tree (multiplied by M's scaling factor) do not violate the
supermartingale property for the composite tree. (If the winning
condition is violated, the corresponding subtree is discarded
and M does not change $t$ inside it anymore.)

\subsubsection*{Trees of variable height}

The correction factors (possible increase of supermartingales in
the game on finite tree) depend on the height of the finite tree
the game is played on. These heights we may choose at our
discretion, and we do this in such a way that the product of A's
coefficients converges while the corresponding product of M's
coefficients diverges.

\subsubsection*{Two winning cases}

Unfortunately, even this scheme (two increase factors depending
on the tree height) is not final. In the actual game on the
finite tree, M wins in any of the two cases:

(1)~either A's supermartingales $t_0$, $t_1$ do not increase at
all on the path from the root to some leaf, whereas M's
supermartingale $t$ in this leaf increases (even a small
increase in enough);

(2)~or A's supermartingales $t_0$, $t_1$ increase, but this
increase is small while M's supermartingale $t$ increases
substantially.

More formally, for a tree of height $h$ there are two pairs of
real numbers, $(M_1(h),m_1(h))$ and $(M_2(h), m_2(h))$. A leaf
is called \emph{winning} (at some step of the game) if the leaf
has the label $i=1$ or
the label $i=2$ (attached by M as explained below)
and $t$ is greater than $M_i(h)$ in the leaf
whereas $t_0$, $t_1$ are not
greater than $m_i(h)$ on the path to the leaf
(here we speak about the current values of the supermartingales).

These pairs are:
        $$
\left(1+\frac{1}{2^h-1},1\right),
\quad
\left(\frac{3}{2},1+\frac{1}{2^{\lfloor (h-1)/2\rfloor}}\right).
        $$
(The exact values are not so important, we need only the
properties mentioned above in~(1) and~(2); in fact, we use only
odd values of $h$.)

As mentioned above, we have one additional requirement: M must
indicate the type of the winning leaf. Namely, M can attach
labels $i=1$ or $i=2$ to leaves; once attached, the label cannot
be removed; a leaf cannot carry both labels. Note that if the
conditions for supermartingales are fulfilled, but the leaf has
no label or ``wrong'' label, this leaf is not a winning leaf.
(We need this to choose correctly the height of the next tree.)

At last, we are ready for an exact statement:

\textbf{Main Lemma on games on finite trees}. \emph{For this
game on a tree of any odd height $h\ge 3$, M has a winning
strategy; this strategy guarantees that at least one winning
leaf appears and remains winning forever}.

We prove the lemma in the next section. Now let us show
how the lemma is used to construct a winning strategy on the
infinite tree.

M starts to play on a tree of a certain height, say $h=3$. When
a winning leaf appears, M adds a new tree on top of this leaf
(having it as a root). This new tree has the same height $h$ if
$i=1$, and it has a larger height $h+2$ if $i=2$. When this
second-level tree gets a winning leaf, M adds a new tree
according to the same rule, and so on. For technical
convenience, we will assume that there is only one winning leaf
at any moment postponing the others candidated until the current
one is rejected. When a leaf is rejected we discard everything
that grows from this leaf.

(Note that Figure~ref{supermart1}
above does not show this adequately: as we
go up, the heights of the trees grow; note also that the heights
of trees of the same level may be different.)

Let us consider the infinite branch (it is unique under our
assumption) obtained in the limit and the labels that appear
along this branch. If there is only a finite number of labels
$i=2$ along it, then the heights of the trees along the branch
are the same from some point, thus A's supermartingales $t_0$,
$t_1$ do not increase after this point and M's supermartingale
$t$ increases every time by a small, but constant factor, hence
$t$ is unbounded.

Now consider the case of infinite number of $i=2$ labels. For
the trees where the winning leaf has the label $i=2$, A's
supermartingales $t_0$, $t_1$ increase at most by the factor
       $$1+1/(2^{\lfloor (h-1)/2\rfloor})$$
for odd~$h$, and each $h$ occurs only once. For the other trees,
$t_0$, $t_1$ do not increase, thus they are bounded. On the
other hand, each of the infinitely many leaves with $i=2$
provides that M's supermartingale $t$ increases by a factor of
$3/2$, and for the other trees, $t$ does not decrease (even
increases a bit), hence $t$ is not bounded. Therefore, in both
cases, we get an infinite branch where A's supermatingales are
bounded, whereas M's supermartingale is not.

\subsubsection*{Winning strategy}

Let us describe M's winning strategy on a tree of some odd
height $h$. Before the game starts, M selects some path from the
root to one of the leaves, for example, the path that always
goes left. Denote the nodes on this path $A_0,A_1,A_2,\ldots$,
and denote their brothers by $B_1,B_2,\ldots$ ($A_0$ denotes the
root).

\begin{figure}[h]
\begin{center}
\includegraphics[scale=1.0]{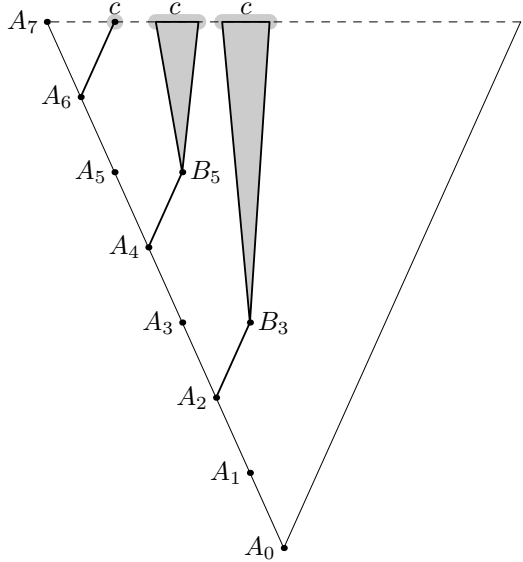}
\end{center}
\caption{The first move of M}\label{supermart3}
\end{figure}

At the first move, M lets the value of supermartingale $t$ to be
$c=2^h/(2^h-1)$ in all the leaves above $B_3,B_5,\ldots$
(Fig.~\ref{supermart3}) and labels all these leafs with $i=1$.
Other leaves keep zero values. The values in the other
(non-leaf) nodes are set so that $t$ is a supermartingale,
i.e.\, as the average over the leaves that are their
descendants. (Recall that initially the value of $t$ is zero
everywhere except the root, where the value is $1$ all the
time.)

As a result of this move, several winning leaves appear. To
avoid defeat, A has (at some point) to increase A's
supermartingales: for each winning leaf, $t_0$ or $t_1$ must be
made greater than $1$ somewhere on the path to this leaf.
Consider two variants: this happens either on the path to one of
$B_3,B_5,\ldots$ or inside the subtrees rooted at
$B_3,B_5,\ldots$

\textbf{First case}:
One of A's supermartingales becomes greater than~$1$ in some
$A_i$ (note that if a node is on the path to $B_j$ then this
node is $A_i$ for some $i$).

\textbf{M's move}:
Set supermartingale $t$ to be equal to $c$ in all leaves except
the leaf on the selected path (the leftmost leaf $A_h$ in our case),
and label all these leaves with $i=1$. Below, $t$ is adjusted by
averaging (thus getting exactly $1$ in the root).

\textbf{Why this helps}: Both A's supermartingales are not
greater than~$1$ in the root. There exists a path where they
both do not increase (let us construct this path from the root:
the supermartingale that does not bet in the node does not
change, and we choose the direction where the betting one does
not increase). Hence, there is a leaf such that on the path to
it both $t_0$, $t_1$ are not greater than~$1$. This cannot be
M's selected path (the leftmost one in our case),
since it goes through all $A_i$, including the ``bad'' one.
But in all other leaves, M's supermartingale $t$ is equal
to $c$, thus one of them is winning.

\textbf{Second case}: In all the subtrees above
$B_3,B_5,\ldots$, A has increased $t_0$ or $t_1$. Therefore, in
all the nodes $B_3,B_5,\ldots$, either $t_0$ or $t_1$ is greater
than~$1$ (otherwise one can find a path from $B_j$ to a leaf
where both $t_0$, $t_1$ are not greater than one, i.\,e.\, a
winning leaf).

Actually, we may assume here that the supermartingale that is
greater than~$1$ in $B_j$ does not bet in $B_j$ and bets in the
previous node $A_{j-1}$: the other possibility is covered by the
first case, since the supermartingale that does not bet in
$A_{j-1}$ would have the same value (greater than~$1$) in
$A_{j-1}$ as in $B_j$. Hence, one supermartingale is greater
than~$1$ in all $B_3,B_5,\ldots$. To be definite, we say that it
is the ``odd'' supermartingale $t_1$ (this is literally true for
first, third, and all the games on finite trees with the odd
numbers, where $t_1$ bets in the roots; for the games with even
numbers, $t_0$ and $t_1$ switch their roles).

\begin{figure}[h]
\begin{center}
\includegraphics[scale=1.0]{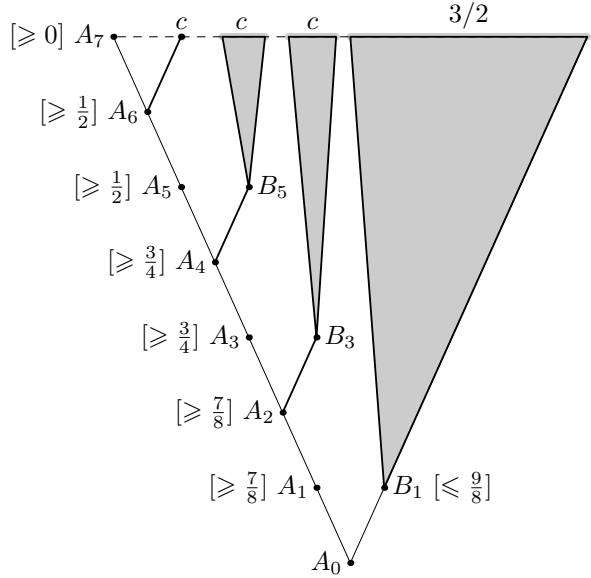}
\end{center}
\caption{Lower bounds for $t_1$ along the M's selected path,
and his second move}\label{supermart4}
\end{figure}

\textbf{Lower bound}.
We can obtain lower bounds for the values of $t_1$ in all $A_i$,
going down along M's selected path. The supermartingale $t_1$
does not bet in half of the nodes (and keeps its value) while in
the other half the lower bound is averaged with a value greater
than~$1$. Thus we get a bound as in Fig.~\ref{supermart4} (the
figure is for the tree of height~$7$).
Since the supermartingale $t_1$ is not greater than~$1$ in the
root and bets there, we get an upper bound for its value in
$B_1$: $9/8$ for the case in the figure and $1+1/2^{(h-1)/2}$ in
general.

\textbf{M's second move}: In all the leaves above $B_1$, the
value of $t$ is set to~$3/2$, and the leaves are labeled
with~$i=2$. Below, $t$ is adjusted by averaging (it is easy to
check that the value in the root does not exceed $1$: in the
second quarter of the leaves, $t$ is zero, and in the first
quarter, $t$ is slightly greater than~$1$ in some leaves only).

\textbf{Why this helps}: In $B_1$, both A's supermartingales are
not greater than $1+1/2^{(h-1)/2}$ (we have shown this for
$t_1$; since $t_0$ does not bet in the root, it does not
exceed~$1$ also in $B_1$), and this holds along some path to a
leaf above $B_1$. But M's supermartingale $t$ is~$3/2$ there.

The construction of the winning strategy for the game on the
finite tree (and thus the proof of the theorem) is completed.

\section{Non-uniform measures}

Recall that martingales and supermartingales can be interpreted
as a capital during a game. So far we have considered a game
where the coin is symmetric. If the next bit is guessed
correctly, the stake is doubled. But we can consider other
distributions where the declared probabilities of $0$ and $1$
are not equal. In this case it is fair that a stake on a less
probable outcome wins more than a stake on a more probable
outcome. These ``non-symmetric games'' correspond to
(super)martingales with respect to a measure. A non-negative
function $\mu$ on binary strings is called a \emph{measure}
(probability distribution) if $\mu(\Lambda)=1$ and
        $$
\mu(x) = \mu(x0)+\mu(x1)
        $$
for all strings~$x$. A \emph{supermartingale with respect to
measure $\mu$} is a non-negative function $m$ on binary strings
that satisfies the inequality
        $$
m(x) \ge m(x0)\frac{\mu(x0)}{\mu(x)}+m(x1)\frac{\mu(x1)}{\mu(x)}
        $$
for all strings~$x$. (For the uniform measure
$\lambda(x)=2^{-length(x)}$, we get our previous definition of
supermartingales.)

When lower semicomputable supermartingales are considered, it is
usually required that the measure is computable. The definition
of Martin-L\"of randomness is extended to the case of arbitrary
computable measures in a natural way, and the randomness
criterion based on lower semicomputable supermartingales works
for the general case too.

We stated our main theorem for supermartingales with respect to
the uniform measure. But it holds for a wider class of
computable measures.

\medskip

\textbf{Theorem.} \emph{Let $\mu$ be a computable measure such that
the conditional probabilities for the next bit are separated from
zero:
        $$
\exists\varepsilon>0\,\forall x\,[\mu(x0)/\mu(x)>\varepsilon
\text{ and } \mu(x1)/\mu(x)>\varepsilon].
        $$
Then there exist a Martin-L\"of non-random with respect to $\mu$
sequence $\omega$ such that no $\mu$-supermartingale that
bets every other step \textup(only on even steps or only on
odd steps\textup) wins on $\omega$.}

\smallskip

The proof follows the same line with minimal changes: one should
adjust thresholds in the winning conditions and select the
path $A_0,A_1,A_2,\ldots$
used in the winning strategy with some caution.

\end{document}


\title{Алгоритмическая случайность и разбиение супермартингалов}
\author{Ан.\,А.\,Мучник\thanks{%
  Работа была выполнена Андреем Альбертовичем Мучником
  (1958--2007) в 2003~году, тогда же он рассказал о ней
  А.\,В.\,Чернову, который подготовил предварительный текст
  статьи. Этот текст был просмотрен Ан.\,А.\,Мучником, который
  предполагал работать над ним дальше, но изменений внести не
  успел. Окончательный текст подготовили А.\,В.\,Чернов и
  А.\,Шень в 2007--2008~годах; указанные лица несут всю
  ответственность за возможные ошибки в тексте. Работа поддержана
  грантами РФФИ и Sycomore (ANR, CNRS, France).}
}
\date{}

\maketitle

\begin{abstract}
Одно из (эквивалентных) определений случайности
последовательности в смысле Мартин-Лёфа использует перечислимые
снизу супермартингалы. Мы показываем, что не всякий такой
супермартингал можно заменить двумя, делающими ставки только на
чётных и только на нечётных членах последовательности.
\end{abstract}

\section[Случайность и перечислимые снизу супермартингалы]
{Случайность и перечислимые снизу\\ супермартингалы}

Одно из определений алгоритмической случайности для бесконечной
последовательности нулей и единиц (относительно бернуллиевского
распределения c независимыми испытаниями и вероятностью успеха
$1/2$) использует так называемые супермартингалы. В
алгоритмической теории вероятностей \emph{супермартингалом}
называется функция, определённая на всех двоичных словах,
принимающая неотрицательные действительные значения и
удовлетворяющая неравенству
        $$
m(x) \ge \frac{m(x0)+m(x1)}{2}
        $$
для любого слова~$x$. Такая функция соответствует стратегии для
игры, в которую мы вступаем с некоторым начальным капиталом
$m(\Lambda)$ (где $\Lambda$~--- пустая строка), и перед каждым
испытанием (бросанием монеты) часть имеющихся на данный момент
денег ставим на нуль, часть на единицу, а часть выбрасываем.
Проигравшая ставка (на неправильный исход) теряется,
а выигравшая возвращается в двойном размере.
В этой терминологии $m(x)$~--- наш капитал после того, как выпали
биты слова~$x$. (Если возможность выбросить часть денег не
используется, то неравенство обращается в равенство, и мы
получаем определение \emph{мартингала}.)

Мы говорим, что (супер)мартингал $m$ \emph{выигрывает} на
последовательности $\omega$, если его значения на начальных
отрезках этой последовательности не ограничены сверху.

В алгоритмической теории вероятностей часто рассматривают
\emph{перечислимые снизу} супермартингалы: это означает, что
$m(x)$ есть предел неубывающей последовательности
неотрицательных рациональных чисел $M(x,0), M(x,1),\ldots$, и
функция $(x,n)\mapsto M(x,n)$ вычислима.

Среди всех перечислимых снизу супермартингалов имеется
наибольший (с точностью до постоянного множителя); его
выигрышное множество (наибольшее по включению среди выигрышных
множеств перечислимых снизу супермартингалов) называется
множеством \emph{неслучайных} последовательностей, а дополнение
этого множества~--- множеством \emph{случайных}
последовательностей.

Это определение эквивалентно классическому определению
Мартин-Лёфа (см., например,~\cite{uss}), поэтому его можно
назвать также критерием случайности по Мартин-Лёфу в терминах
супермартингалов.

\textbf{Замечание}. Иногда рассматривают также и большее
множество последовательностей, которое получится, если в этом
определении заменить перечислимые снизу супермартингалы на
вычислимые мартингалы (для простоты будем считать, что их
значения~--- рациональные числа); среди вычислимых мартингалов
нет наибольшего, так что последовательность называют случайной,
если \textsl{все} вычислимые мартингалы на ней ограничены.

\section{Чётные и нечётные шаги}

Для вычислимых мартингалов имеет место такое простое наблюдение:
при определении случайности можно ограничиться
мартингалами, которые делают ставки либо только на чётных
шагах, либо только на нечётных шагах. Будем говорить, что
(супер)мартингал \emph{делает ставки только на чётных шагах},
если $m(x)=m(x0)=m(x1)$ для всех слов $x$ чётной длины.
(Например, отказ от ставки на третьем шаге означает, что после
третьего бросания капитал не меняется, то есть
$m(x0)=m(x1)=m(x)$ при двухбитовых~$x$.)

Аналогичным образом (супер)мартингал \emph{делает ставки только на
нечётных шагах}, если $m(x)=m(x0)=m(x1)$ для всех слов нечётной
длины. (Говоря об отсутствии ставки, мы имеем в виду, что
капитал делится поровну между нулём и единицей, и мы в любом
случае остаёмся <<при своих>>.)

Сформулируем точно упомянутое выше наблюдение: \emph{для любого
вычислимого мартингала $t$ существуют два вычислимых мартингала
$t_0$ и $t_1$, первый из которых делает ставки только на чётных
шагах, а второй~--- на нечётных, и выполнено такое свойство:
если $t$ выигрывает на некоторой последовательности $\omega$, то
хотя бы один из мартингалов $t_0$ и $t_1$ выигрывает на
$\omega$.} Отсюда следует, что выигрышное множество мартингала
$t$ содержится в объединении выигрышных множеств мартингалов
$t_0$ и $t_1$.

\textbf{Доказательство}: добавив константу, можно считать, что
мартингал $t$ всюду положителен. Построим два мартингала $t_0$ и
$t_1$ так: на чётных шагах $t_0$ делит капитал между нулём и
единицей в той же пропорции, что $t$, а $t_1$ не делает ставок
(=ставит поровну на нуль и единицу), а на нечётных шагах~---
наоборот. Тогда выигрыш $t$ (в процентах от начального капитала)
равен произведению выигрышей $t_0$ и $t_1$, так что если эти два
мартингала ограничены на начальных отрезках какой-то
последовательности $\omega$, то и $t$ ограничен на них.

\medskip

Другими словами можно сказать так: при определении случайности
относительно вычислимых мартингалов можно ограничиться
мартингалами, которые делают ставки только через раз (либо
только на чётных шагах, либо только на нечётных). Аналогичное
замечание (также почти очевидное) можно сделать и для
определения случайности по Мизесу -- Чёрчу (с помощью правил
выбора, см.~\cite{uss}): достаточно разбить правило выбора на
два, одно выбирает только члены подпоследовательности с чётными
индексами, другое~--- только с нечётными.

Основной результат работы состоит в том, что для перечислимых
снизу супермартингалов (и случайных по Мартин-Лёфу
последовательностей) аналогичное утверждение неверно.

\medskip

\textbf{Теорема.} \emph{Существует неслучайная по Мартин-Лёфу
последовательность $\omega$, на которой не выигрывает ни один
перечислимый снизу супермартингал, делающий ставки только на
чётных шагах, а также ни один перечислимый снизу супермартингал,
делающий ставки только на нечётных шагах}.

\smallskip

Доказательство этого результата будет дано в следующих разделах
статьи. Сейчас мы отметим его связь с теоремой Ламбальгена о
случайности пары.

Каждую последовательность нулей и единиц $\alpha$ можно разбить
на две подпоследовательности (чётные и нечётные члены) и тем
самым сопоставить с ней пару последовательностей $\alpha_0$ и
$\alpha_1$. Несложно понять, что для случайной (в смысле
Мартин-Лёфа) последовательности $\alpha$ обе эти
последовательности ($\alpha_0$ и $\alpha_1$) будут случайны,
однако это условие хотя и необходимо, но не достаточно
(очевидный контрпример: $\alpha_0=\alpha_1$).

Известная теорема М.\,Ламбальгена~\cite{lam} говорит, что мы
получим необходимое и достаточное условие, если потребуем
случайности $\alpha_0$ с оракулом $\alpha_1$ и наоборот.
Неформально говоря, это соответствует тому, что игрок имеет
возможность просматривать (любые) члены последовательности
$\alpha_1$, делая ставки на $\alpha_0$, и наоборот. Можно было
бы предположить, что нет необходимости в <<просмотре
вперёд>> (когда мы, делая ставку на некоторый нечётный бит,
просматриваем чётные, \emph{идущие после него}, и наоборот)~---
но, как показывает сформулированная теорема, это не так.

Получается парадокс: представим себе, что двое судей бросают
монеты перед началом игр какого-то турнира, но делают это по
очереди (один по чётным дням, другой по нечётным. Было бы
естественно ожидать, что если каждый из них делает это
добросовестно (соответствующая подпоследовательность случайна в
контексте всей последовательности, см.~\cite{online}), то и вся
последовательность будет случайна. Оказывается, что если
определять случайность с помощью супермартингалов, то это не
так: результирующая последовательность может быть неслучайной,
но вину за это нельзя возложить ни на одного из судей.

\section{Построение игры, соответствующей теореме}

Доказательство теоремы удобно изложить в виде некоторой
бесконечной игры двух лиц, которые будут именоваться <<мы>>
и <<противник>>.

Противник постепенно строит два супермартингала $t_0$ и $t_1$
(делающие ставки на чётных и нечётных шагах соответственно); мы
строим один супермартингал $t$. Построение состоит в том, что на
каждом шаге игроки указывают некоторое приближение снизу к этим
супермартингалам, которое имеет лишь конечное число ненулевых
значений и само является супермартингалом требуемого вида; в
процессе игры эти супермартингалы могут только расти (игроки не
могут уменьшить уже заявленные значения). Каждый ход любого из
игроков, таким образом, можно считать конструктивным объектом
(закодировать двоичным словом и т.\,п.); игроки ходят по очереди
и видят ходы друг друга (игра с полной информацией).
Дополнительно мы требуем, чтобы значения всех супермартингалов на
пустом слове равнялись единице (нормировка, необходимая для
того, чтобы значения не уходили в бесконечность).

Партия представляет собой бесконечную последовательность
чередующихся ходов; в результате партии возникает три предельных
супермартингала. Считается, что мы выиграли в такой
(бесконечной) партии, если существует последовательность
$\omega$, на которой наш супермартингал $t$ выигрывает (не
ограничен на начальных отрезках $\omega$), в то время как оба
супермартингала противника $t_0$ и $t_1$ не выигрывают
(ограничены на начальных отрезках).

\textbf{Основная лемма}. \emph{Существует вычислимая стратегия в
этой игре, гарантирующая нам выигрыш.}

Несложно заметить, что этой леммы достаточно для доказательства
теоремы. В самом деле, стандартным образом можно доказать, что
существуют два максимальных (с точностью до ограниченного
множителя) перечислимых снизу супермартингала $t_0$ и $t_1$
описанного вида (делающих ставки только на чётных/нечётных
шагах). Запустим стратегию, использовав в качестве противника
перечисление снизу этих двух супермартингалов. Поскольку
стратегия вычислима, то и даваемый ей супермартингал $t$ будет
перечислим снизу. Поскольку стратегия выигрывающая, то найдётся
последовательность $\omega$, на которой $t$ выигрывает, а $t_0$
и $t_1$ (и, значит, любые перечислимые снизу супермартингалы
указанного вида) не выигрывают, откуда и вытекает утверждение
теоремы.

Осталось доказать лемму. Это будет сделано в следующих двух
разделах: мы определим версии этой игры на конечном дереве и
предъявим стратегию, гарантирующую выигрыш в этих версиях (в
некотором точном смысле), а затем объясним, как из этих конечных
стратегий собрать стратегию для бесконечной игры.

\section{Игры на конечных и бесконечных деревьях}

Рассмотренную нами игру удобно представлять себе на бесконечном
дереве, вершинами которого являются двоичные слова. Игроки
увеличивают оценки своих супермартингалов для вершин этого
дерева, а после окончания бесконечной партии (если можно так
выразиться) приходит арбитр и ищет бесконечную ветвь с
требуемыми в определении игры свойствами.

Мы будем рассматривать также похожие игры на конечных деревьях.
Полем такой игры является полное двоичное дерево некоторой
высоты $h$. Как и раньше, игроки ходят по очереди, увеличивая
текущие значения своих супермартингалов ($t$ для нас, $t_0$ и
$t_1$ для противника). Условия на $t_0$ и $t_1$ те же: $t_0$
может делать ставки только на чётных шагах, $t_1$~--- только на
нечётных. Значения супермартингалов (всех трёх) в корне всегда
равны единице; в остальных вершинах изначально супермартингалы
равны нулю и постепенно увеличиваются игроками.

Игра по-прежнему бесконечна, хотя число вершин и конечно; для
данной (бесконечной) партии мы рассматриваем предельные значения
всех трёх супермартингалов в листьях дерева и формулируем условие
выигрыша в их терминах. Это условие по техническим причинам
довольно сложное, но смысл его такой: мы хотим, чтобы в одном из
листьев дерева наш супермартингал $t$ был заметно больше~$1$, в
то время как на пути к этому листу оба супермартингала
противника не превосходили бы единицы, или в крайнем случае
очень ненамного превосходили единицу
(по сути, это неасимптотический вариант условия на бесконечном
дереве, требущего, чтобы $t$ стремилось к бесконечности, а
$t_0$, $t_1$ были ограничены).

Представим себе на минуту, что нам удалось этого достичь в
следующей (более сильной, чем мы сможем на самом деле, но об
этом потом) форме: в одном из листьев наш супермартингал больше
$2$, а оба супермартингала противника по-прежнему меньше~$1$. Такую
стратегию можно использовать как составную часть при построении
выигрышной стратегии на бесконечном дереве: над выигрышным
листом мы можем построить новое такое же поддерево, где начать
ту же самую игру, но уже имея двукратную фору. (Это значит, что
ходы в игре на бесконечном дереве вдвое больше ходов на конечном
дереве второй игры, и что мы применяем выигрышную стратегию для
этого конечного дерева, умножая ходы на два.) В пределе получим
лист, где наш супермартингал больше $4=2^2$, а оба супермартингала
противника по-прежнему не больше~$1$. На нём можно вырастить ещё
одну копию дерева, где получить уже $2^3=8$, и так далее~--- и в
пределе получится бесконечная ветвь, где наш супермартингал
стремится к бесконечности, а супермартингалы противника ограничены.

Это описание является упрощённым и игнорирует некоторые моменты.
Прежде всего надо иметь в виду, что игра на конечном дереве не
заканчивается, поэтому мы должны начинать следующие игры (на
присоединённых поддеревьях), не дожидаясь её конца и будучи
готовыми к тому, что выбранный лист впоследствии окажется не
выигрышным. Из нашего описания выигрыша в игре на конечном
дереве видно, что условие выигрыша в данном листе может сначала
выполниться (наш супермартингал станет больш\'им, а у противника
оба супермартингала пока малы), а потом нарушиться (когда
супермартингал противника увеличится), но после этого уже вновь
выполниться не сможет~--- так что повторно ни один лист в
кандидаты не попадает.

Выпуская поддерево из текущего кандидата (и бракуя его и всех
потомков, когда кандидат будет отброшен), мы получаем картину
вроде следующей (рис.~\ref{supermart1}).
\begin{figure}[h]
\begin{center}
\includegraphics[scale=1.0]{supermart1.mps}
\end{center}
\caption{Активные и отменённые игры на конечных деревьях}\label{supermart1}
\end{figure}
Cерые треугольники изображают поддеревья, где игра была
запущена, а потом оказалось, что зря (кандидат под ними был
отменён); светлые треугольники изображают действующие (на
некоторый момент) игры. Точками изображены кандидаты, в которых (на
данный момент) выигрышная ситуация. (Мы будем считать, что в
каждый момент в каждом поддереве выбирается какой-то один из
возможных кандидатов, над которым строится поддерево следующей
игры, и пока этот кандидат не отброшен, над другими деревья не
строятся. В принципе можно было бы строить деревья следующего
уровня над всеми кандидатами.)

Заметим, что на самом деле нам нужно играть против одного
противника на бесконечном дереве (а не против разных
противников на конечных поддеревьях). Однако ходы противника на
бесконечном дереве естественно переводятся в ходы в играх на
(активных в данный момент) конечных деревьях.

По определению выигрыша на каждом следующем уровне наш
супермартингал возрастает в два раза по сравнению с его
значением на предыдущем уровне; с другой стороны, по тому же
определению вдоль всего пути, проходящего по выигрышным листьям,
супермартингалы противника не превосходят~$1$.

Посмотрим, что произойдёт в пределе с этой картинкой. На первом
уровне возможно только конечное число изменений кандидатов,
поэтому в некоторый момент появится ``окончательный кандидат'',
который никогда не отменится. С этого момента игра над ним будет
активной, в ней тоже будет только конечное число изменений и так
далее. В итоге мы получим бесконечную ветвь, на которой наш
супермартингал неограничен, а оба супермартингала противника
ограничены.

\section{Игра на конечном дереве}

\subsubsection*{Идея конструкции}

Начнём с неформального объяснения ``причин'', по которым в игре
на конечном дереве существует выигрышная стратегия. Читатель,
которому эти объяснения покажутся непонятными, может сразу
перейти к
точным формулировкам и доказательствам в следующих разделах,
которые на эти объяснения не опираются.

Начнём с очевидных замечаний. Мы хотим, чтобы в некотором листе
дерева наш супермартингал~$t$ был больше~$1$. Это означает, что
в каких-то других листьях он должен быть меньше~$1$ (поскольку
среднее значение супермартингала в листьях не превосходит значения в
корне, которое равно~$1$). Таким образом, нам нужно как-то найти
листья, на которые можно не тратить денег попусту, тем самым
сэкономив деньги для других листьев. Одна из возможностей~--- не
тратить деньги на ``бесперспективные'' листья, на
пути к которым один из супермартингалов противника
уже превысил единицу.
Если мы поставим некоторое $C>1$ во все остальные листья,
кроме этого бесперспективного, то это гарантирует нам выигрыш:
на некотором пути из корня оба супермартингала
противника не превосходят единицы (в каждой вершине один из
супермартингалов не играет, а второй при одном из исходов проигрывает),
и путь этот закончится в одном из листьев, где мы поставили $C$.
(Чтобы не превысить среднее значение
для нашего супермартингала в корне, можно взять $C=N/(N-1)$, где
$N$~--- число листьев.)

Что же делать, если противник не совершает таких ошибок и не
создаёт бесперспективных листьев, в которых мы ещё
ничего не поставили? Начнём с того, что поместим $C$
в самый левый лист $x$.

\begin{figure}[h]
\begin{center}
\includegraphics[scale=1.0]{supermart5.mps}
\end{center}
\caption{Начало игры. В левой колонке указан
супермартингал, играющий на данном уровне.}\label{supermart5}
\end{figure}

Противник должен рано или поздно на это
ответить, сделав один из своих
супермартингалов $t_0,t_1$ б\'ольшим
единицы в $x$ или на пути к нему. Заметим, что во втором
случае он уже совершит ошибку, о которой мы говорили (появится
бесперспективный лист~$y$~--- брат~$x$).

Поэтому мы можем считать, что один из супермартингалов
противника больше единицы в сам\'ом листе~$x$. Более того, можно
считать, что таковым оказался
супермартингал, который делает ставки в отце~$x$
(обозначим эту вершину $u$);
пусть, скажем, это будет~$t_0$. (В противном случае значение
этого супермартингала в $u$ было бы больше единицы и $y$
оказался бы бесперспективным, хотя мы туда ещё ничего
не поставили.)

Нашим следующим
ходом мы помещаем $C$ в~$y$.
В ответ противник должен увеличить на пути в $y$
один из своих супермартингалов. Это не может быть $t_0$,
поскольку тогда появятся бесперспективные листья. (Если $t_0$ превышает $1$
в $y$, то и в $u$ есть превышение, поскольку в $u$ происходит
усреднение, а в $x$ превышение уже было. Если есть превышение
в $u$, то есть превышение и в его отце
$v$, поскольку в $v$ супермартингал $t_0$
ставок не делает.)

Таким образом, в некоторый момент супермартингал $t_1$
превысит $1$ в вершине $y$ или на пути к ней. Поскольку на
последнем шаге супермартингал  $t_1$ не играет, то это превышение
произойдёт уже в $u$ или на пути к  $u$.

После этого мы помещаем $C$ в следующий (третий слева)
лист~$z$; противник должен сделать один из
супермартингалов большим
единицы в $z$ или на пути к $z$, и это может быть только
супермартингал $t_0$, иначе следующий лист~$p$ станет
бесперспективным. Затем мы помещаем $C$ в $p$;
один из супермартингалов противника должен превысить
единицу на пути в~$p$, и это может быть только $t_0$,
иначе $t_1$ превысит единицу в $u$ и $q$ (отце $p$),
а следовательно,
и в $v$ (усреднение), а потому и в предшествующей
вершине $w$ (где $t_1$ не играет),
и появятся сразу четыре бесперспективных листа.

Продолжая этот процесс, мы видим, что на каждом шаге у
противника есть только одно возможное действие, если он хочет
избежать преждевременной отбраковки вершин, и потому мы знаем,
какой из его
супермартингалов должен превысить единицу в каждом из
листьев (кроме последнего). Эту закономерность можно пояснить
аналогией. Представим себе, что в каждом листе имеется булева
переменная, значение которой ($0$ или $1$) определяет, какой из
супермартингалов
противника ($t_0$ или $t_1$) превышает единицу в данном листе.
Значения этой переменной распространяются вниз (к корню) по
простым И-ИЛИ-правилам: если супермартингал играет в вершине и
его значение превосходит единицу в \emph{обоих} сыновьях этой
вершины, то он превосходит единицу и в этой вершине; если же он
не играет в вершине, то достаточно превосходить единицу в
\emph{одном} из сыновей.

Преждевременная бесперспективность листа
соответствует при этой аналогии тому, что называется сокращением
вычислений (short-cut evaluation) в языках программирования:
если одна из частей конъюнкции ложна (или одна из частей
дизъюнкции истинна), отпадает надобность в вычислении другой.
Противник, присваивающий значения переменным в листьях
И-ИЛИ-дерева слева направо, и желающий, чтобы значение каждой
следующей переменной (при известных предыдущих) могло повлиять
на значение выражения, также не имеет выбора (до последнего
момента).

Идея выигрышной стратегии: либо мы пользуемся преимуществом,
которое даёт нам преждевременно дискредитированный лист (нам не
надо тратить на него деньги), либо используем преимущества,
которые возникают после того, как противник обслужит первую
четверть всех листьев. В первом случае мы сможем поставить
$C=N/(N-1)$ на все листья, кроме одного (преждевременно
дискредитированного); здесь $N$ обозначает число листьев. Во
втором случае посмотрим на последовательность булевских
значений, при которой не происходит сокращения вычислений, то
есть $0100010101000100\ldots$\footnote{%
        Эта последовательность рассматривается в
        http://www.research.att.com/njas/sequences/A035263 и
        http://www.research.att.com/njas/sequences/A096268.}
Каждый второй член в этой последовательности нулевой, то есть в
каждом втором листе супермартингал $t_0$ превышает единицу.
Поэтому двумя уровнями ниже все значения $t_0$ не меньше $1/2$,
и подчиняются тому же закону, так что в половине из них $t_0$ не
меньше~$1$. Ещё двумя уровнями ниже все значения $t_0$ не меньше
$3/4$, а половина не меньше~$1$, и так далее. Аналогичное
рассуждение применимо к $t_1$ со сдвигом на уровень. Поэтому мы
можем взять любую вершину, над которой достаточно большое
поддерево, и рассчитывать на то, что в ней (скажем) $t_1$ если и
меньше~$1$, то ненамного. Возьмём в качестве этой вершины самого
левого из четырёх внуков корня (в котором $t_1$ делает ставки; в
детях корня он их не делает). Тогда $t_1$ будет только чуть-чуть
меньше $1$ и в левом сыне корня.
А значит, $t_1$ не может быть сильно больше
$1$ в правом сыне корня, и если мы сосредоточим остатки денег на
листьях правой половины дерева (на каждый лист придётся примерно
$4/3$, поскольку четверть листьев остаются свободными и ещё в
одной четверти в них стоит $C$). С другой стороны, оба
супермартингала противника лишь немного могут превосходить $1$ в
правом сыне корня, и это свойство распространяется снизу вверх
хотя бы в один лист.

В обоих случаях мы достигаем некоторого преимущества над
противником, но преимущества эти различны, и точное определение
конечной игры должно это учитывать. Перейдём к формальному
описанию этой игры (которое не зависит от изложенных выше
неформальных соображений).

\subsection*{Игра на конечном дереве: точное определение}

Как мы упоминали, сформулированная нами для примера цель игры
(добиться, чтобы наш супермартингал был равен $2$ в вершине, на
пути к которой супермартингалы противника не превышают~$1$)
сформулирована слишком сильно: мы сможем достичь меньшего, но
этого окажется достаточно.

\subsubsection*{Разрешённый рост для супермартингалов}

В игре на конечном дереве нам не удастся гарантировать, что в
одном из листьев (и на пути к нему) оба супермартингала
противника будут в точности не больше~$1$; некоторый их рост (в
$1+\delta$ раз) придётся разрешить. Эти коэффициенты
($1+\delta$) перемножаются для деревьев цепочки и накапливаются,
но если произведение их ограничено константой, это не страшно.

С другой стороны, нам не удастся и гарантировать увеличение
нашего субмартингала в $2$ раза, а лишь в некоторое меньшее
число раз (на каждом уровне, как мы увидим, своё). В этом случае
нам, напротив, необходимо, чтобы произведение этих коэффициентов
стремилось к бесконечности.

Говоря более формально, для каждого из конечных поддеревьев мы
применяем лемму с соответствующим поправочными множителями для
значения супермартингалов в корне. Таких множителей два: для нас
и для противника, и они перемножаются отдельно. Условие выигрыша
в предыдущих листьях гарантирует, что реальные ходы противника
не превысят разрешаемых в игре на конечном дереве (с поправкой),
и что наши ходы не нарушат условие супермартингальности. (Как
только один из листьев на пути перестаёт быть выигрышным, мы
прекращаем изменения в поддереве.)

\subsubsection*{Деревья переменного размера}

Сказанное об изменении коэффициентов возрастания (для нашего
супермартингала и для супермартингалов противника) предполагает,
естественно, что высоты конечных деревьев, на которых ведутся
игры, различны. Эти высоты мы можем выбирать по своему усмотрению,
стремясь достичь нашей цели (расходимости произведения наших
коэффициентов и сходимости произведения коэффициентов
противника).

\subsubsection*{Двухвариантный выигрыш}

К сожалению, и эта картина (два коэффициента увеличения~--- для
нас и для противника,~--- зависящие от высоты) является
упрощённой. На самом деле у нас будет два варианта выигрыша; для
данной высоты $h$ дерева будут две пары чисел $(M_1(h),m_1(h))$
и $(M_2(h), m_2(h))$. Выигрышным листом считается такой, в
котором (для некоторого $i\in\{1,2\}$) наш супермартингал
превосходит $M_i(h)$, в то время как оба супермартингала
противника в этом листе и на пути к нему не превосходят
$m_i(h)$.

Эти пары будут такими:
        $$
\left(1+\frac{1}{2^h-1},1\right),
\quad
\left(\frac{3}{2},1+\frac{1}{2^{\lfloor (h-1)/2\rfloor}}\right).
        $$
Конкретный вид этих формул не так важен. Существенно то, что для
выигрыша нам достаточно существования листа, в котором выполнено
одно из двух: либо

(1)~наш супермартингал умножается (по сравнению с корнем) на
некоторый коэффициент, хотя бы ненамного больший~$1$, в то время
как супермартингалы противника не увеличиваются, либо

(2)~наш супермартингал существенно увеличивается (в $3/2$ раза) ценой
небольшого (и быстро убывающего с ростом~$h$) увеличения у
противника.

При этом есть ещё одно уточнение: мы обязаны указывать, какой
вид выигрыша мы планируем в данном листе. Это означает, что в
ходе игры мы можем ставить на листьях (нестираемые и
несовместные друг с другом) пометки ``1'' и ``2'' (вид
выигрыша), и должны обеспечить, чтобы в пределе существовал лист
с пометкой, в котором достигается указанный тип выигрыша.
(Пометка на листе определяет высоту следующего дерева, растущего
из этого листа.)

Итак, мы готовы к окончательной формулировке:

\textbf{Основная лемма об играх на конечных деревьях}. \emph{Для
любого нечётного $h\ge 3$ в игре на дереве высоты $h$ у нас есть
стратегия, гарантирующая один из двух указанных видов выигрыша}.

Покажем, как из этой леммы вытекает существование выигрышной
стратегии на бесконечном дереве. (Доказательство леммы мы
отложим до следующего раздела).

В самом деле, начнём играть на дереве некоторой высоты,
например, $3$. Если у него появляется выигрышный лист, то над
ним мы строим новое дерево~--- той же высоты $3$, если лист
первого рода, и большей высоты, скажем, $5$, если лист второго
рода. Если у этого дерева второго уровня появляется выигрышный
лист, то мы строим над ним следующее дерево~--- не меняя высоту,
если лист первого рода, и увеличивая высоту на $2$, если второго
рода. (Как мы уже говорили, можно считать, что в каждый момент
выигрышный лист только один, откладывая использование следующего
до тех пор, пока предыдущий окажется негодным. Когда лист
становится негодным, всё построенное над ним также
отбрасывается.)

(Отметим, что приведённый выше рисунок~\ref{supermart1}
 становится неадекватным:
высота деревьев растёт снизу вверх, а не убывает, как
нарисовано, и, кроме того, высота разных деревьев одного уровня
может быть разной.)

Посмотрим, что может получиться в пределе на бесконечной ветви.
В ней либо конечное число выигрышных листьев второго рода, либо
бесконечное. Если конечное, то высоты деревьев в этой ветви
стабилизируются, супермартингалы противника с некоторого момента
не растут (и потому ограничены), а наш супермартингал с этого же
момента умножается на постоянный коэффициент (и потому
неограничен). Если бесконечное, то при каждом переходе второго
рода супермартингалы противника умножаются на
       $$1+1/(2^{\lfloor (h-1)/2\rfloor})$$
с очередным~$h$, а в промежутках не растут, так что в целом
последовательность органичена. С другой стороны, каждый переход
второго рода увеличивает наш супермартингал в $3/2$ раза, а в
промежутках он не уменьшается (и даже немного увеличивается, но
это не важно), так что в целом он не ограничен.

Таким образом, в обоих случаях мы получаем бесконечную ветвь,
где супермартингалы противника ограничены, а наш~--- нет.

\subsubsection*{Выигрышная стратегия: подробности}

Опишем выигрышную стратегию на дереве нечётной высоты $h$.
Выберем путь из корня в один из листьев, например, путь, идущий
всё время налево. Вершины вдоль этого пути обозначим
$A_0,A_1,A_2,\ldots$. Мы начнём игру с того, что рассмотрим
братьев $B_3,B_5,\ldots$ и в деревьях над ними напишем число
$c=2^h/(2^h-1)$ (рис.~\ref{supermart3}). Под ними мы тоже должны
увеличить значение нашего супермартингала, положив его в каждой
вершине среднему значению в листьях-потомках. (Напомним, что
изначально все значения нашего супермартингала равны нулю, кроме
единицы в корне, которая так и остаётся единицей на протяжении
игры.)
\begin{figure}[h]
\begin{center}
\includegraphics[scale=1.0]{supermart3.mps}
\end{center}
\caption{Наш первый ход}\label{supermart3}
\end{figure}

В результате этого появляется сразу несколько вершин, где наш
супермартингал не меньше~$c$. Противник, чтобы избежать
проигрыша (по первому варианту), должен сделать один из своих
супермартингалов больше единицы на пути в любой из листьев, в
которых стоит~$c$. Он может делать это либо в деревьях над
$B_3,B_5,\ldots$, либо ниже их. Начнём со второго варианта.

\textbf{Случай~А}. Противник сделал один из своих
супермартингалов больше~$1$ в одной из вершин $A_i$ (только эти
вершины встречаются на пути к $B_i$ не доходя до них).

\textbf{Наши действия}: увеличиваем наш супермартингал во всех
вершинах, кроме самой левой, до $c$. (В остальных вершинах
производим усреднение; в корне это усреднение даёт как раз
единицу.)

\textbf{Почему этого достаточно}: поскольку оба супермартингала
противника в корне не больше~$1$, то найдётся лист, на пути к
которому они оба не больше~$1$ (ищем лист из корня вверх, в
каждой вершине один из супермартингалов делает ставку, идём
вверх так, чтобы он не возрос~--- а второй заведомо не
изменится). Этот лист не может быть самым левым (на пути к нему
один из супермартингалов противника бывает больше единицы), а в
остальных у нас стоит~$c$.

\textbf{Случай~Б}. Противник увеличил свои супермартингалы в
деревьях над $B_3,B_5,\ldots$, закрыв все вершины, где мы
поставили~$c$. В этом случае во всех вершинах $B_3,B_5,\ldots$
один из супермартингалов противника больше единицы. (В самом
деле, иначе из этой вершины можно было бы протянуть путь вверх,
вдоль которого оба супермартингала противника не превосходят
единицы.)

\textbf{Нижняя оценка}. Можно считать без ограничения общности,
что превосходит единицу тот из супермартингалов, который перед
этим играл (поскольку другой супермартингал не меняется, другой
вариант покрывается случаем А). Таким образом, можно считать,
что ``нечётный'' супермартингал противника больше~$1$ во всех
вершинах $B_3,B_5,\ldots$

В этом случае можно получить нижнюю оценку на значения нечётного
супермартингала в вершинах $A_i$, идя сверху вниз. В половине
случаев он не делает ставок и потому не изменяется; в оставшейся
половине мы производим усреднение с числом, большим $1$.
Получаем такую оценку, как на рисунке~\ref{supermart4} (где она
показана для дерева высоты~$7$):
\begin{figure}[h]
\begin{center}
\includegraphics[scale=1.0]{supermart4.mps}
\end{center}
\caption{Нижние оценки для $t_1$ вдоль избранного пути и второй
ход}\label{supermart4}
\end{figure}
Поскольку в корне нечётный супермартингал не больше~$1$, то
в $B_1$ он не больше $9/8$ для нашего рисунка,
а в общем случае~--- не больше $1+1/2^{(h-1)/2}$).

\textbf{Наши действия}: во всех листьях над $B_1$ мы полагаем
наш супермартингал равным~$3/2$, ниже по усреднению (легко
проверить, что мы не превысим единицы в корне, так как вторая
четверть листьев не использована вообще, а в первой четверти
значение лишь чуть больше~$1$ и не во всех вершинах).

\textbf{Почему этого достаточно}: так как в $B_1$ оба
супермартингала противника не больше $1+1/2^{(h-1)/2}$ (а один
даже не превосходит единицы, но это неважно), то это же верно на
пути к одному из листьев над $B_1$. А наш супермартингал в любом
таком листе равен~$3/2$.

Заметим, что требование декларировать тип выигрыша выполняется
легко: вершины с~$c$ мы декларируем как вершины первого типа, а
вершины с $3/2$ декларируем как вершины второго типа.

Построение выигрышной стратегии в игре на конечном дереве (и тем
самым доказательство теоремы) завершено.

\section{О неравномерных мерах}

Супермартингалы соответствуют изменению капитала в ходе игры. До
сих пор мы предполагали, что игра использует симметричную
монету, и тогда ставка на угаданный бит удваивается. Если монета
предполагается несимметричной, то и правила игры должны
соответственно измениться: менее вероятный исход даёт большее
увеличение ставки. Это отражено в определении супермартингала
относительно заданной меры. Неотрицательная функция~$\mu$ на
двоичных словах называется (вероятностной) \emph{мерой}, если
$\mu(\Lambda)=1$ (где $\Lambda$~--- пустое слово) и
        $$
\mu(x)=\mu(x0)+\mu(x1)
        $$
для любого $x$. \emph{Супермартингалом относительно}~$\mu$
называется неотрицательная функция $m$ на двоичных словах, для
которой
        $$
m(x) \ge m(x0)\frac{\mu(x0)}{\mu(x)}+
         m(x1)\frac{\mu(x1)}{\mu(x)}
        $$
при всех $x$. (Для равномерной меры, где $\lambda(x)=2^{-n}$ для
слова $x$ длины~$n$, получаем прежнее определение.)

Когда говорят о перечислимых снизу супермартингалах, обычно
предполагают, что мера~$\mu$ вычислима. В этом случае сохраняет
силу критерий случайности по Мартин-Лёфу в терминах перечислимых
снизу супермартингалов.

Основной результат этой статьи был сформулирован для случая
равномерной меры, но он переносится и на более широкий класс
вычислимых мер:

\textbf{Теорема}. \emph{Пусть $\mu$~--- вычислимая мера, для
которой условные вероятности появления нуля и единицы после
любого слова $x$ отделены от нуля, то есть не меньше некоторого
положительного $\varepsilon$, не зависящего от~$x$. Тогда
найдётся последовательность, не случайная по мере~$\mu$, на
которой ограничен любой перечислимый снизу супермартингал,
делающий ставки только на чётных или только на нечётных ходах.}

Доказательство следует той же схеме, только надо изменить
численные значения параметров выигрыша на конечном дереве, а
также подходящим образом выбирать путь $A_0,A_1,A_2,\ldots$
для выигрышной стратегии.